\documentclass[11pt]{article}
\usepackage{latexsym}

\usepackage{amssymb}
\usepackage{amsmath}

 \usepackage{epsfig}
 \usepackage{graphicx}

\usepackage{color}

\newcommand{\be}{\begin{equation}}
\newcommand{\ee}{\end{equation}}
\newcommand{\ba}{\begin{eqnarray}}
\newcommand{\ea}{\end{eqnarray}}
\newcommand{\bi}{\begin{itemize}}
\newcommand{\ei}{\end{itemize}}
\newcommand{\baa}{\begin{array}}
\newcommand{\eaa}{\end{array}}
\newcommand{\edoc}{\end{document}}

\newcommand{\nn}{\nonumber \\}
\newcommand{\nr}[1]{(\ref{#1})}
\newcommand{\la}[1]{\label{#1}}

\newcommand{\rmi}[1]{{\mbox{\scriptsize #1}}}

\newcommand{\fra}[2]{\textstyle{\frac{#1}{#2}\,}}  

\newcommand{\bfx}{{\bf x}}

\newcommand{\lambdalow}{\Lambda_\rmi{T}}
\newcommand{\lambdahigh}{\Lambda_\rmi{ET}}
\newcommand{\lambdaETC}{\Lambda_\rmi{ETC}}
\newcommand{\Tlow}{T_\rmi{T}}
\newcommand{\Thigh}{T_\rmi{ET}}

\def\CL{{\cal L}}

\def\gsim{\raise0.3ex\hbox{$>$\kern-0.75em\raise-1.1ex\hbox{$\sim$}}}
\def\lsim{\,\,\raise0.3ex\hbox{$<$\kern-0.75em\raise-1.1ex\hbox{$\sim$}}\,\,}

\begin{document}

\begin{titlepage}
\begin{flushright}
CP3-Origins-2010-14\\
HIP-2010-10/TH\\
\today\\ 
\end{flushright}
\begin{centering}
\vfill

{\Large{\bf Thermodynamics of Quasi Conformal Theories From Gauge/Gravity Duality}}

\vspace{0.8cm}

\renewcommand{\thefootnote}{\fnsymbol{footnote}}

J. Alanen$^{\rm a,b}$\footnote{janne.alanen@helsinki.fi},
K. Kajantie$^{\rm a,b}$\footnote{keijo.kajantie@helsinki.fi}
K. Tuominen$^{\rm b,c}$\footnote{kimmo.tuominen@jyu.fi}\footnote{On leave of absence from Department of
Physics, University of Jyv\"askyl\"a}
\setcounter{footnote}{0}

\vspace{0.8cm}

{\em $^{\rm a}$%
Department of Physics, P.O.Box 64, FI-00014 University of Helsinki,
Finland\\}
{\em $^{\rm b}$%
Helsinki Institute of Physics, P.O.Box 64, FI-00014 University of
Helsinki, Finland\\}
{\em $^{\rm c}$%
CP$^3$-Origins, Campusvej 55, 5230 Odense, Denmark\\}
\vspace*{0.8cm}

\end{centering}

\noindent
We use gauge/gravity duality to
study the thermodynamics of a generic almost conformal theory, specified by its
beta function. Three different phases are identified, a high temperature phase of
massless
partons, an intermediate quasi-conformal phase and a low temperature confining phase.
The limit of a theory with infrared fixed point, in which the
coupling does not run to infinity, is also studied.
The transitions between the phases are of first order or
continuous, depending on the parameters of the beta function.
The results presented follow from gauge/gravity duality; no specific boundary
theory is assumed, only its beta function.

\vfill \noindent


%

\vspace*{1cm}

\noindent


\vfill

\end{titlepage}

\section{Introduction}
Gauge/gravity duality has been proposed to describe various aspects of SU($N_c$)
gauge theories. The studies on thermodynamics \cite{kiri3,kiri4,aks,ak}
have mostly concentrated on QCD due to the existence of experimental
data on ultrarelativistic heavy ion collisions and the existence of
numerical lattice data \cite{boyd,teper,panero}. However, it is plausible that
at larger energy scales other gauge theories may play a significant role. A prime
example is Technicolor (TC) \cite{TC} (for a review, see \cite{Sannino:2009za}),
which effectively replaces the fundamental scalar Higgs field by a $\bar Q Q$ composite.

The main purpose of this article is to lay a framework for the study of the thermodynamics
of generic walking Technicolor-related theories within the framework of
gauge/gravity duality. When the Standard Model (SM) degrees of freedom are included, 
this will, for example, be relevant for the expansion of the Universe through the electroweak phase
transition. There is extensive literature on this in the framework of the Standard Model but
only a limited amount within TC theories, on the thermal aspects
\cite{Nussinov:1985xr,3japs,Jarvinen:2009pk} and on relic dark matter
\cite{Gudnason:2006yj,Kainulainen:2006wq}.

Of course, the details of the TC theory are
unknown. A great advantage of the gauge/gravity duality approach is that this is
not needed, duality is only used to compute expectation values in the boundary theory.
This is also concretely manifest in applications of duality
to condensed matter physics (see, for example, \cite{zaanen}).
The information on the boundary theory we will need is a knowledge of the beta
function of its gauge coupling. This will contain a number of parameters which
can be obtained if the dynamics of the underlying theory is known.

A general property of TC theories we shall assume is that they be of walking type. This assumes
that there are two widely different energy scales, $\lambdalow$ and
 $\lambdahigh\sim 10^3\lambdalow$, between
which the coupling constant of the theory evolves very slowly, the theory is
almost conformal. Below $\lambdalow$ and above $\lambdahigh$ the coupling runs similarly
to asymptotically free theories (like QCD).
In (Extended) Technicolor theories $\lambdalow=\Lambda_\rmi{TC}\approx246$ GeV and
$\lambdahigh=\Lambda_\rmi{ETC}$.
The need for walking behavior comes
from the requirement that the successes of the Standard Model should not be spoiled: the
contributions from new physics to the electroweak precision parameters should be small and
the contributions to flavor changing neutral current interactions should be suppressed.

Concrete examples of walking TC theories with minimal matter content have been suggested in
\cite{sanninotuominen,Dietrich:2005jn}. These have also been studied on the lattice
\cite{Catterall:2007yx,DelDebbio:2008wb,Catterall:2008qk,Hietanen:2008mr,
Hietanen:2009az, Shamir:2008pb,DelDebbio:2009fd,Fodor:2009ar,Bursa:2009tj,
Sinclair:2009ec}. Of course, it could be that the origin of flavor patterns and fermion masses is
not Extended Technicolor (ETC) but some other extension of simple TC dynamics at higher energy scale \cite
{Simmons:1988fu,Antola:2009wq,Dine:1990jd,Antola:2010nt}.

The beta-function ansatz we shall use is
\be
\beta(\lambda)=-c\lambda^2{(1-\lambda)^2+e\over 1+a\lambda^3},\qquad
\lambda=N_cg^2,
\la{betafn}
\ee
which is tuned to asymptotic freedom in the UV ($\lambda\to0$) and to walking near $\lambda=1$
if $e$ is small.
A plot can be found in Fig.~\ref{betasene0}.
The values of parameters $c,a,e$ will be described in detail later. The case $e=0$
(for a plot, see Fig.~\ref{betase0}) will play
a special role: the theory then has an infrared fixed point (IRFP), already studied in
this model in \cite{ak}.

Using methods in \cite{kiri3,kiri4,aks,ak} we derive
the intuitively viable and interesting
finite temperature phase
diagram shown later in Fig.~\ref{phaseD}.
Namely, associated with the changes
in the evolution of the coupling constant at $\lambdahigh$ and
$\lambdalow$ we will find phase transitions with critical
temperatures $\Thigh$ and $\Tlow$, respectively.
Above $\Thigh$ we have
deconfined partonic plasma consisting of (techni)quarks and
(techni)gluons while below $\Tlow$
we have confined (techni)hadronic matter.
Between these temperatures and stretching over a wide range,
$\Tlow\ll \Thigh$, we have a novel new phase of matter,
{\em{quasi-conformal matter}}. This terminology is appropriate since over
the extent of scales spanned by this phase we have $\beta(\lambda)\simeq 0$,
and physics is almost conformal. For $e\not=0$ the evolution never ends in the
infrared fixed point and at least a small
amount of conformality breaking will be always present.

Our approach here is 5 dimensional phenomenological bottom-up and the boundary theory is
in a thermal state. Running couplings of the
walking type at $T=0$ have also been studied top-down starting from 10 dimensional
supergravity solutions in \cite{piai1,piai2}.

A beta function, of course, is scheme dependent. The consequences
we derive, an equation of state and associated phase structure, are
entirely physical. One can thus say that our model defines the
regularisation scheme leading to the coupling constant and its beta
function we start from.

The paper is organized so that in Sec. \ref{gravity} we describe
the gauge/gravity setup which we use.
Our main analysis begins in Sec. \ref{IRFP} with
a special case of the model beta function corresponding
to a theory featuring a stable infrared fixed point, and
in Sec. \ref{walkingbeta} we carry out the analysis for walking technicolor. In Sec. \ref{disc} we will
discuss how our results can be interpreted within the phenomenological contexts of walking TC
and ETC or unparticles, Sec.~\ref{checkout} contains conclusions.

\section{The gravity dual}
\label{gravity}
\subsection{Equations}
The gravity equations of the model are as follows \cite{kiri3,kiri4,aks,ak}.
The model starts from a metric ansatz
\be
ds^2=b^2(z)\left[-f(z)dt^2+d\bfx^2+{dz^2\over f(z)}\right]
\la{ansatz}
\ee
plus a scalar field $\phi(z)=\log\lambda(z)$. The three functions
$b(z), f(z)$ in the metric and the scalar field $\phi(z)$
in the gravity action (in standard notation)
\be
S={1\over16\pi G_5}\int d^5x\,\sqrt{-g}\left[R-\fra43(\partial_\mu\phi)^2+V(\phi)\right]
\la{Eframeaction}
\ee
are determined from the three equations ($\dot b\equiv b'(z)$, etc.)
\ba
&&6{\dot b^2\over b^2}+3{\ddot b\over b}+3{\dot b\over b}{\dot f\over f}={b^2\over f}V(\phi),
\label{eq1}\\
&& 6{\dot b^2\over b^2}-3{\ddot b\over b}={\fra43} \dot\phi^2,\label{eq2}\\
&&{\ddot f\over \dot f}+3{\dot b\over b}=0.\label{eq3}
\ea
Further, from the functions so evaluated, the beta function follows as
\be
\beta(\lambda)=b{d\lambda\over db},\quad \lambda(z)  = e^{\phi(z)}\sim  g^2N_c.
\label{crucial}
\ee
Thus $\lambda(z)$ is the coupling and $b(z)$ is its the energy scale. Note that the
equation for the scalar field follows algebraically from \nr{eq1}-\nr{eq3}.

We are interested in solutions which are asymptotically ($z\to0$) AdS$_5$, i.e.,
$b(z\to0)=\CL/z$
which have a zero at some $z$, $f(z_h)=0$. These have
an entropy and Hawking temperature $4\pi T=-\dot f(z_h)$ and their field theory dual
will be the thermal system we are searching for. When $b=\CL/z$, \nr{eq1} implies
$V(z=0)=12/\CL^2,\,f(0)=1$. Furthermore, we will always consider theories asymptotically
free in the UV, i.e., $\beta(\lambda\to0)=-c\lambda^2$. Asymptotic freedom, asymptotic
AdS$_5$ and \nr{crucial} together imply $dz/z=d\lambda/(-c\lambda^2)$, i.e.,
\be
\lambda(z\to0)={1\over c\log(1/\Lambda z)},
\la{lasmallz}
\ee
where $\Lambda$ is a constant of integration of Eq.\nr{crucial}. One can, of course,
add a 2-loop term in the beta function, but this will effectively just change $\Lambda$
to a 2-loop $\Lambda$. Note that \nr{lasmallz} implies $\phi(z\to0)=-\log\log(1/z)\to-\infty$.

The constant $\Lambda$ can also be related to the normalisation of $b$ as follows.
By integrating \nr{crucial} one has
\be
\log{b\over b_0}=\int_{\lambda_0}^\lambda{d\lambda\over\beta(\lambda)}\to\log{\CL\over
b_0z}\approx
{1\over c\lambda}
\ee
which together with \nr{lasmallz} implies that
\be
\Lambda={b_0\over\CL}.
\ee

To solve Eqs. (\ref{eq1})-({\ref{eq3})
one simplifies by introducing
\be
W= -\dot b/b^2.
\la{defW}
\ee
One ends up with the system of equations
\ba
\dot W&=& 4bW^2-\fra{1}{f}(W\dot f+\fra13 b V),\nn
\dot b&=& -b^2W,\nn
\dot \lambda&=&\fra32\lambda\sqrt{b\dot W},\nn
\ddot f&=&3\dot fbW,
\la{system}
\ea
which one can proceed to solve numerically when $V$ is known.
We do not know $V$, we only know the beta function. However, we can get
a candidate $V$ by first solving from \nr{eq2} and \nr{crucial}
\be
W(\lambda)=W(0)\exp\left(-\fra49\int_0^\lambda d\bar\lambda{\beta(\bar\lambda)
\over\bar\lambda^2}\right),\quad W(0)={1\over\CL}.
\la{W}
\ee
Since the beta function is a property of
the vacuum theory, $f=1$, we can now find
a candidate
potential simply by inserting \nr{W} and \nr
{defW}
to \nr{eq1} with $f=1$:
\be
V(\lambda)=12W^2(\lambda)\left[1-\left({\beta\over3\lambda}\right)^2\right].
\la{V}
\ee
Apart from a slowly varying logarithmic term (see later Eq.~\ref{Vansatz}),
this is the potential we shall use and see numerically that the solutions
reproduce very accurately the beta function we started from.
For black hole solutions $\beta(\lambda)$ as computed from \nr{crucial} will show
thermal effetcs in the infrared, clearly, for example, $\lambda<\lambda(z_h)$
(see later Fig.~\ref{betasene0}) .

Expanding in the limit $\lambda\to0$ with $\beta(\lambda)=-c\lambda^2$ one finds that
$V(\lambda=e^\phi)=12/\CL^2(1+8c\lambda/9+..)$. This contrasts with many other scalar
+ gravity
models having, at small $z$, $V=12/\CL^2 -\fra12 m^2\phi^2+..$.

As an example, consider our model beta function \nr{betafn}. Integration of \nr{W}
gives
\ba
\log[W(\lambda)/W(0)]&=& \fra{2c}{27a}
\left[2 \sqrt{3} a^{1/3} (-2 +
        a^{1/3} (1 + e)) (\arctan[{-1 + 2 a^{1/3} \lambda\over \sqrt{3}}] -
        \arctan{-1\over\sqrt{3}}) + \right.\nn&& \left.\hspace{-3cm}
     a^{1/3} (2 + a^{1/3} (1 + e)) \log[(1 + a^{1/3} \lambda)^2/
     (1 - a^{1/3} \lambda + a^{2/3} \lambda^2)] +
     2 \log[1 + a \lambda^3])\right]\nn
     &\equiv& w(\lambda;c,a,e).
\la{wmodel}
\ea
The limits of this rather untransparent expression near $\lambda=0,1,\infty$ are
\ba
\log[W(\lambda)/W(0)]&=& \fra{4c}{9}
\left[(1+e)\lambda-\lambda^2+\fra13\lambda^3+...\right]\la{smallla}\\
&=&\log w(1;c,a,e)+\fra{4ce}{9(1+a)}(\lambda-1)+.. \\
&=&
\fra{4c}{81a}\left[9\log(a^{1/3}\lambda)+2\pi\sqrt{3}
((1+e)a^{1/3}-2)+\fra{18}{\lambda}-\fra{9(1+e)}{2\lambda^2}+...\right]\la{largela}
\ea
The general pattern is shown in Fig.\ref{logwfig}: linear increase at small $\lambda$
and a transition to the large-$\lambda$ behavior
$
W(\lambda)={\rm const}\,\cdot\,\lambda^{4c/(9a)},\quad V(\lambda)={\rm const}\,\cdot
\,\lambda^{8c/(9a)}
$ around $\lambda\sim1$.

\begin{figure}
\begin{center}
\includegraphics[width=.5\textwidth]{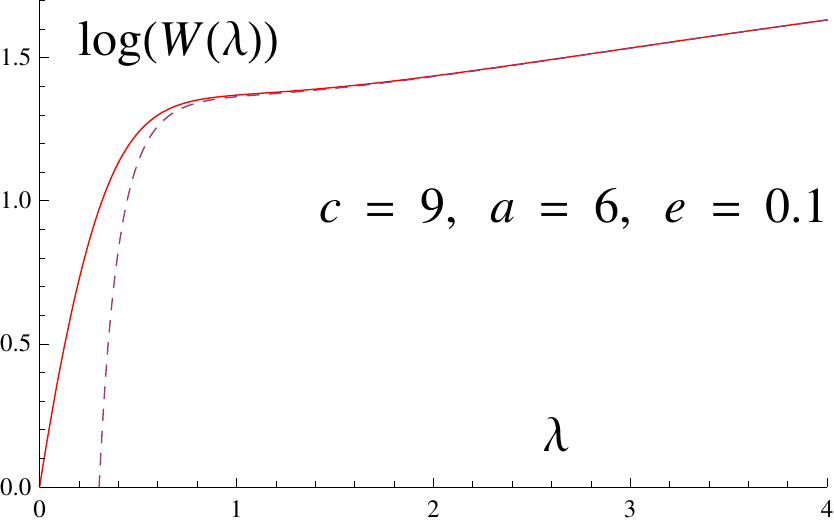}
\caption{The function $W(\lambda)$ for $c=9,\,a=6,\,e=0.1$, also $\CL=1$. The dashed line is the
large-$\lambda$ approximation \nr{largela}.}
\label{logwfig}
\end{center}
\end{figure}

\subsection{Numerical integration}

Numerical integration of \nr{system} cannot start at $z=0$ or at $z_h,\,f(z_h)=0$. Instead, it
is convenient to start at some initial $z=z_i=z_h-\varepsilon$, $\varepsilon$ = some small
number.
Computing analytically one finds that the initial values of various functions are
\ba
\lambda_i&=&\lambda_h-\fra38\lambda_h^2b_h^2{V'(\lambda_h)\over -\dot f_h}\,\varepsilon,
\nn
b_i&=&b_h+b_h^2W_h\,\varepsilon,\nn
W_i&=&W_h-\fra{1}{16\dot f_h^2}b_h^3\lambda_h^2(V'(\lambda_h))^2\,\varepsilon,\nn
f_i&=&f_h-\dot f_h\,\varepsilon,\nn
\dot f_i&=&\dot f_h-3\dot f_h b_hW_h\,\varepsilon,
\la{initcond}
\ea
where $\lambda_i\equiv\lambda(z_i)$, $\lambda_h\equiv\lambda(z_h)$, similarly for the
others.
Among the five initial quantities at $z_h$ in \nr{initcond} one firstly has $f_h=0$ by definition,
the regularity of the $1/f$ term in \nr{system} requires
\be
W_h={b_hV(\lambda_h)\over3(-\dot f_h)},
\ee
the value $b_h$ will be traded for the constant of integration $\Lambda$ in \nr{lasmallz}, the
value of $\dot f_h$ will be traded for $T$ and, finally, $\lambda_h$ parametrises the different
solutions.

Requiring that $-\dot f_h>0$ and $V'(\lambda_h)>0$ one sees that $\lambda_i$ and $W_i$
are less
than their values at horizon, the others are greater. This is as it should be since $\lambda(z),
\,W(z)$
are monotonically increasing and $b(z),\,f(z)$ monotonically decreasing functions.

Assume now we have a set of values of $\lambda_h,\,b_h,\,-\dot f_h$. For each of these a
numerical integration
of \nr{system} produces a solution $\lambda(z),W(z),b(z),f(z)$ over some range
$z_m<z<z_h$, $b(z_m)$
diverges. This solution is now processed as follows:
\begin{enumerate}
\item Scale $W(z_m)$ to the value one. Define a scaling factor $S_1=W(z_m)$ and write
$\lambda_1(z)=\lambda(z),\,W_1(z)=W(z)/S_1,\,b_1(z)=S_1b(z),\,f_1(z)=S_1^2f(z)$. This
also
scales $f_1(z_m)=1$.

\item Shift $z_m$ to zero. Define $S_2=z_m$ and write
$\lambda_2(z)=\lambda_1(z+z_m),\,W_2(z)=W_1(z+z_m),\,b_2(z)=b_1(z+z_m),\,f_2(z)=f_1
(z
+z_m)$.

\item Scale $z$ so that for each solution, for any $\lambda_h$, Eq.\nr{lasmallz} holds with
some given $\Lambda$. The reason is that this is the only place where the solution is known
analytically so that one use it to fix constants. We know that, for small $z$,
\be
\lambda_2(z)={1\over-c\log(\Lambda_2 z)}={1\over-c\log(\Lambda(\Lambda_2z/\Lambda))}.
\ee
Thus one sees that the proper scaling is $S_3=\Lambda_2/\Lambda$, leading to the
new solution
$\lambda_3(z)=\lambda_2(z/S_3),\,W_3(z)=W_2(z/S_3),\,b_3(z)=b_2(z/S_3)/S_3,\,f_3(z)
=f_2
(z/S_3)$.
Here $\Lambda_2$ is the constant $\exp(-1/c\lambda_2(z))/z$.
This set with index 3 is the final solution. Note that the value of $\lambda_h$ has remained
unchanged and thus parametrises the solution.

\end{enumerate}

Thermodynamics can now be constructed as discussed in \cite{kiri4}. One chooses a numerical
value for $\Lambda$ (we used $\Lambda=1/200$), a small UV value for $\lambda(z)$ where
\nr{lasmallz} is valid (we used $\lambda_\rmi{UV}=0.02$), a set of values for $\lambda_h$
(we used $0.021<\lambda_h\lsim100$, for $e=0$ only values up to $1$ are needed), integrates
the equations as described above, calculates from the solution $T=T(\lambda_h)$ and $s=s(\lambda_h(T))$
and integrates $p=p(T)$ from $p'(T)=s(T)$. All the thermodynamics is then obtained from $p(T)$ and
its derivatives. Concrete expressions to be evaluated are in Eqs.~(22)-(26) of \cite{ak}.

\begin{figure}[!tb]
\begin{center}

\includegraphics[width=0.46\textwidth]{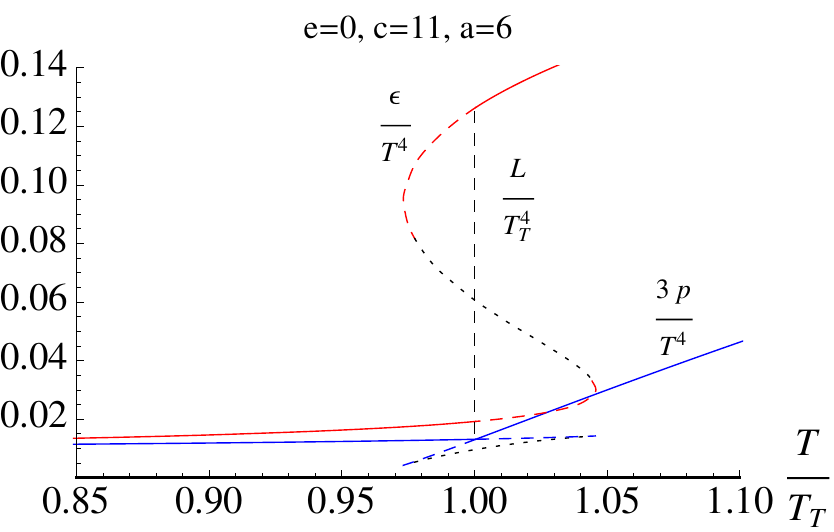}
\hfill
\includegraphics[width=0.46\textwidth]{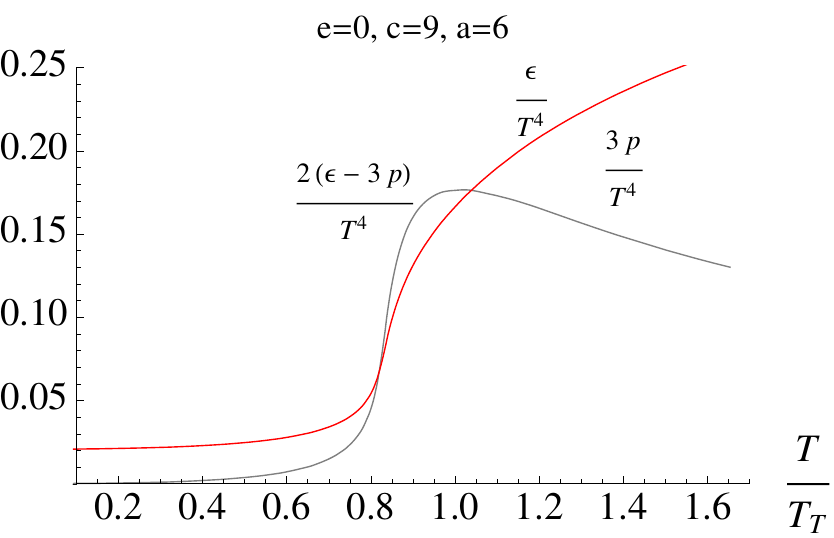}

\end{center}

\caption{\small Energy density and pressure/$T^4$ for $e=0$ for a first order (left panel)
and for a continuous transition (right panel). For a continuous transition also the interaction
measure is given (multiplied by 2 for clarity); $T_T$ then is defined as the location of its maximum.
Parameter values are shown in the figure. The numbers needed to apply \nr{dofs} to relate the
$T\to\infty$ and $T\to0$ limits are $w(1;11,6,0)=3.998$, $w(1;9,6,0)=3.108$.
}
\la{eose0}
\end{figure}

\begin{figure}[!tb]
\begin{center}

\includegraphics[width=0.46\textwidth]{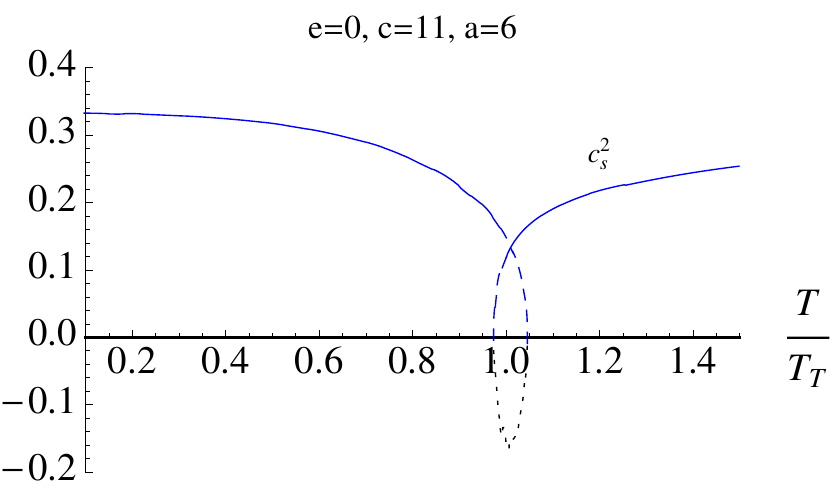}
\hfill
\includegraphics[width=0.46\textwidth]{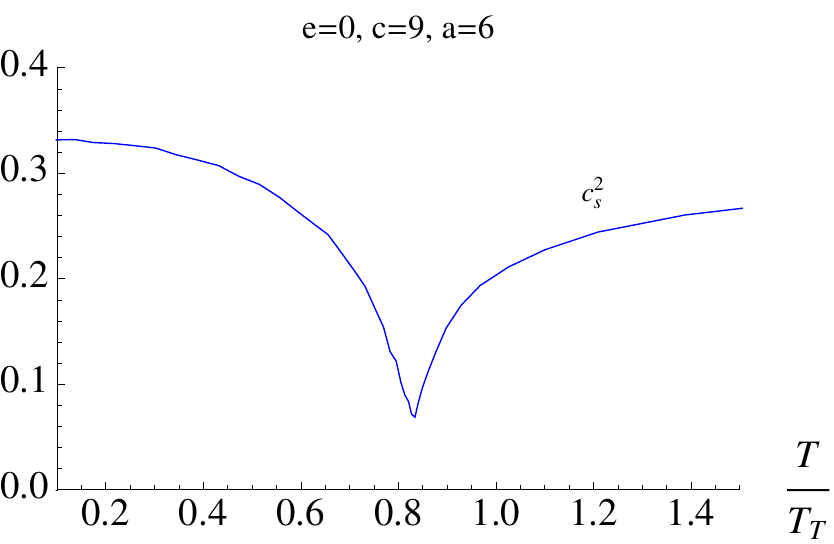}

\end{center}

\caption{\small Sound speed squared for $e=0$ for a first order (left panel)
and for a continuous transition (right panel). For 1st order transition the dashed
segments correspond to the supercooled and -heated branches in Fig.\ref{eose0},
the dotted segment to the unstable branch.
Parameter values are shown in the figure. For a cross-over the dip in $c_s^2$ and maximum
of interaction measure need not coincide.
}
\la{cse0}
\end{figure}

\begin{figure}[!tb]
\begin{center}

\includegraphics[width=0.46\textwidth]{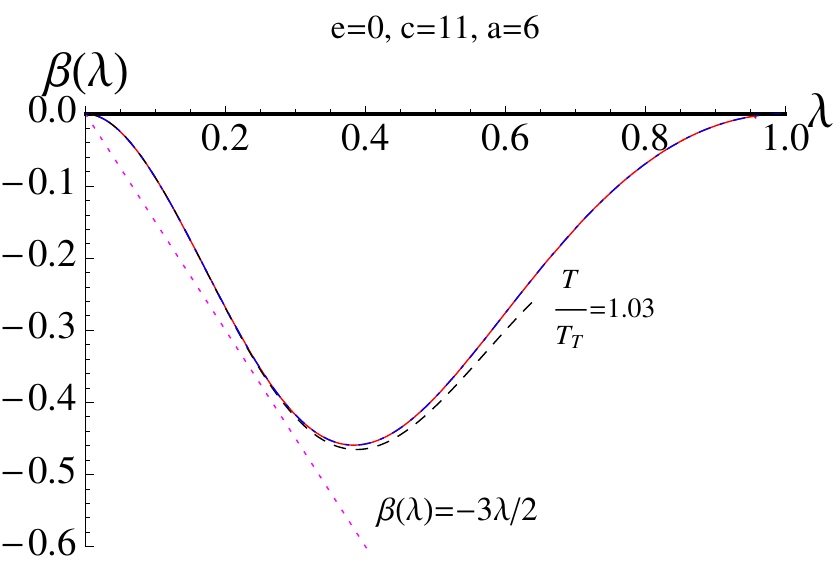}
\hfill
\includegraphics[width=0.46\textwidth]{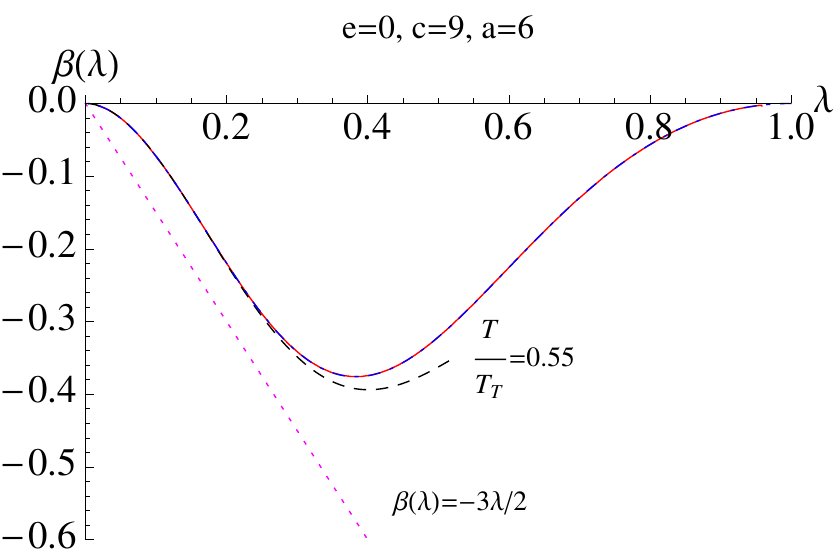}

\end{center}

\caption{\small The input beta function (parametrisation \nr{betafn} with $e=0$, continuous curve)
and the output beta function, computed by inserting to \nr{crucial} the numerical solution
for $T=0.55\Tlow$ or $T=1.03\Tlow$(black dashed curve) and for the
smallest value of $T$ in the computation, $T\approx 0$ (dotdashed curve),
for parameter values shown in the figure. The continuous and dotdashed
curves coincide to the accuracy of the figure.
The confinement line \cite{kiri2} $-\fra32\lambda$ is also shown.
}
\la{betase0}
\end{figure}

\section{Infrared fixed point}
\label{IRFP}
We begin with taking $e=0$ in (\ref{betafn}) and
study the case where an infrared stable FP (IRFP) exists
at $\lambda=1$. The dilaton potential is given by \nr{V},
with $W$ as given by \nr{wmodel} with $e=0$. Thermodynamics for this case has already been
studied in \cite{ak} using a simpler beta function $\beta(\lambda)=-c\lambda^2 (1-\lambda)$.
The approach towards $\lambda=1$ is somewhat different, but
the resulting thermodynamics is nevertheless qualitatively similar.

The coupling now runs from $0$ in the UV to $1$ in the IR.
The theory is conformal in both ends and, accordingly, the transition is between black hole
states in two asymptotically AdS$_5$ spaces with radii $\CL_\rmi{UV}$ and $\CL_\rmi{IR}$ defined by
\be
{1\over\CL_\rmi{UV}}=w(0;c,a,0),\qquad{1\over\CL_\rmi{IR}}=w(1;c,a,0),
\ee
where $w(\lambda;c,a,e)$ is given by \nr{wmodel}. Due to the normalisation in \nr{wmodel},
\be
{\CL_\rmi{UV}\over\CL_\rmi{IR}}=w(1;c,a,0)>1.
\la{Ls}
\ee
The value of the parameter $a$ is inessential now that $\lambda$ is bounded, we take
$a\approx2c/3$ as later in the unbounded $e>0$ case. The crucial parameter is $c$:
the transition is of first order for large $c$ and weakens when $c$ is decreased.
Below some threshold value the transition turns to a continuous one. The threshold value is
approximately determined on whether the beta function crosses the confinement line
$-\fra32\lambda$; see Fig.~\ref{betase0}.

Examples of the resulting equations of state and of the sound velocities squared
$c_s^2=dp/d\epsilon$ are shown in Figs. \ref{eose0} and \ref{cse0}.
For $c=11$ one has a first order transition between a "gluonic" phase and between an
"unparticle" phase, for $c=9$ the transition is a cross-over, for $c\approx9.95$ a 2nd order one. 
At $T\to\infty$ the model implies \cite{ak}
\be
{p\over T^4}\to{\CL_\rmi{UV}^3\over4G_5}{\pi^3\over4}
\la{pT4}
\ee
and the numerics is so normalised that
\be
{p\over T^4}\to {\pi^2\over45}.
\ee
The curves can simply be scaled to proper normalisation $p/T^4=g_\rmi{eff}\pi^2/90$ if $g_\rmi{eff}$
is known. However, what is important is that \nr{pT4} and \nr{Ls} now imply that
\be
\lim_{T\to0}{p\over T^4}={1\over w^3(1,c,a,0)}\lim_{T\to\infty}{p\over T^4},
\la{dofs}
\ee
the effective number of degrees of freedom in the "unparticle" phase is determined
by that in the "gluonic" phase and is less. The relevant numbers are $w(1;11,6,0)=3.998$,
$w(1;9,6,0)=3.108$ and the explicit computations in Fig.~\ref{eose0}
are seen to be in agreement with \nr{dofs}.

Since the potential $V(\lambda)$ was derived analytically from the
$T=0$ field theory beta function \nr{betafn},
it is of interest to compare this input beta function with that one obtains
by inserting numerically computed solutions for $\lambda(z),\,b(z)$ to \nr{crucial}.
These define a temperature dependent beta function. Fig.~\ref{betase0} shows the
pattern: with decreasing $T$ the $T$-dependent beta function approximates the input
beta function stepwise better and better until at $T\to0$ the input beta function
is obtained. The whole scheme is thus entirely consistent.

\begin{figure}[!tb]
\begin{center}

\includegraphics[width=0.6\textwidth]{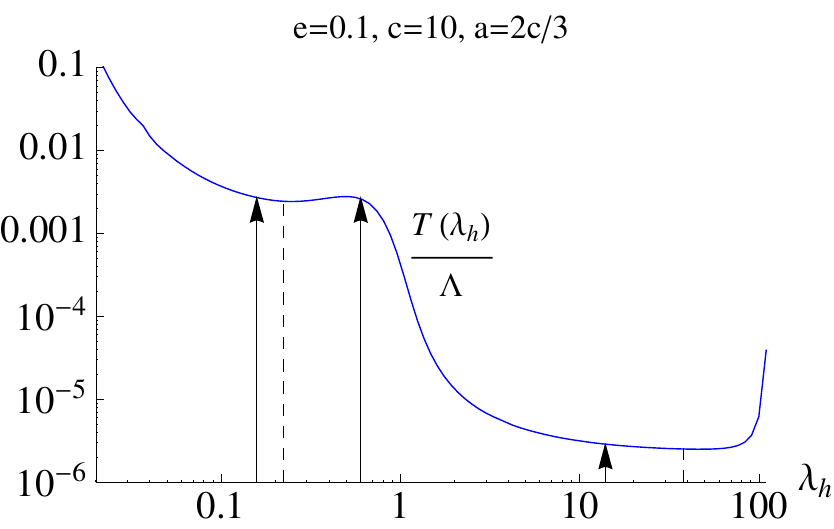}
\end{center}

\caption{\small The temperature in units of $\Lambda$ computed for $c=3a/2=10,\,e=0.1$.
The minima (dashed lines) are at $\lambda_h=0.248$ and $\lambda_h=44.9$. Crucial values are the two
$\lambda_h$'s which have the same $T$ and pressure but are in different branches of
decreasing $T$: these are $\lambda_h=0.173$ and $=0.617$. Finally, the quasiconformal
$\to$ confining transition happens at $\lambda_h=13.81$: there $p$ goes to zero.
}
\la{Tlahfig}
\end{figure}

\begin{figure}[!tb]
\begin{center}

\includegraphics[width=0.8\textwidth]{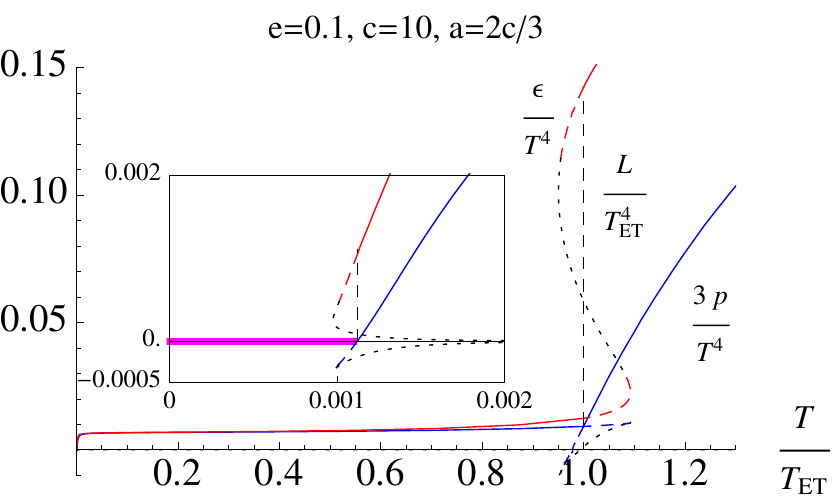}
\end{center}

\caption{\small Energy density and $3p$ scaled by $T^4$ plotted vs $T/\Thigh$ for the beta
function \nr{betafn} with $e=0.1,\,c=3a/2=10$. The normalisation is such that at $T\gg \Thigh$
both $\epsilon/T^4$ and $3p/T^4$ approach $\pi^2/15$.
There is a first order transition at $T=\Thigh$,
metastable branches are dashed and the unstable branch is dotted. There is a second 1st order phase
transition at $T=\Tlow\approx 0.0011\Thigh$, the details are shown in the inset. In between,
for $\Tlow<T<\Thigh$ there is a quasiconformal phase with nearly constant $p/T^4$. The thick
line below $\Tlow$ corresponds to the non-black hole low $T$ phase with $p=0$.
}
\la{eos_eneq0}
\end{figure}

\section{Walking technicolor beta function}
\label{walkingbeta}

Consider now the beta function \nr{betafn} for small but nonzero $e$.
The potential $V(\lambda)$ is again evaluated from \nr{V} with $W$ given by \nr{wmodel}, 
but an additional modification is needed \cite{kiri3}.

To fix the parameters we first note that, to model walking and quasiconformality, $e$ inherently
must be some small number. We shall start with $e=0.1$ and study the effects of varying $e$
at the end of this Section. Next, we shall choose $c/a=3/2$. The motivation for this is that,
as discussed in \cite{kiri2}, a condition for
confinement is that the equation
\be
\beta(\lambda)+\fra32\lambda=0
\la{confeq}
\ee
have a solution. The existence of a $\bar QQ$ condensate is a requisite for any
TC model and at low $T$ confinement typically implies also the formation of
a condensate.  Thus we need confinement and 
\nr{confeq} should have a solution at large $\lambda$. If $c/a=3/2$, our beta function
at large $\lambda$ is parallel to the line $\beta=-\fra32\lambda$ and \nr{confeq}
does not have a solution at finite $\lambda$. However, as shown in 
\cite{kiri2}, a solution is obtained if the potential \nr{V} is
modified by a logarithmic factor so that, at large $\lambda$,
$V\sim\lambda^{4/3}(\log\lambda)^P$. At large $\lambda$ 
this effectively multiplies the beta function by
a term $1+3P/(4\log\lambda)$ and \nr{confeq} has a solution. 
If $P=\fra12$ one obtains
a glueball spectrum with $M^2$ linear in a discrete index $n$.
Though we now do not know what the analogue of the glueball spectrum would be,
we nevertheless stick to this property and take $V$ to be
\be
V=12W^2(\lambda)\left[1-\beta^2/(9\lambda^2)\right]
\left[1+\fra{e}{10}\sqrt{\log(1+\lambda^4)}\right],
\label{Vansatz}
\ee
with $W$ computed from \nr{W} and $\beta$ in \nr{betafn}. The power $\lambda^4$ within
the log is chosen so that it does not affect the leading small-$\lambda$ behavior.

Thermodynamics computed from here for $c=3a/2=10$, $e=0.1$
is shown in Figs.~\ref{Tlahfig}-\ref{betasene0}.
One again has a first order transition, and even two of them, for large $c$.
Crucial for this is that $T(\lambda_h)$ have two minima, as shown explicitly
in Fig.~\ref{Tlahfig}. We remind that $\lambda_h$
parametrises numerical solutions of the gravity equations \nr{eq1}-\nr{eq3}.
Stable phases correpond to $dT/d\lambda_h<0$ and with the structure in
Fig.~\ref{Tlahfig} one can have $T$ equal in two different decreasing branches
of $T(\lambda_h)$, the precise location is determined by also pressure being the
same in the two branches. Decreasing $T$ from very large values one obtains a
first order transition at $T=\Thigh$ with the 1st order structure shown in
Fig.~\ref{eos_eneq0}. Below $\Thigh$ there follows a long quasiconformal
phase in which $p/T^4$ is almost constant. For the IRFP case in the previous
Section this extended down to $T=0$; now at about $T\approx0.01\Thigh$ pressure
starts decreasing and crosses $p=0$ at some $T=\Tlow=0.0011\Thigh$. Below this
the quasi conformal phase becomes metastable (see inset in Fig.~\ref{eos_eneq0})
and the vacuum phase with $p=0$ is the stable one. This is again a first
order transition from a quasiconformal phase to a confining phase, the energy density
$\epsilon$ dropping suddenly to zero.

\begin{figure}[!b]
\begin{center}

\includegraphics[width=0.6\textwidth]{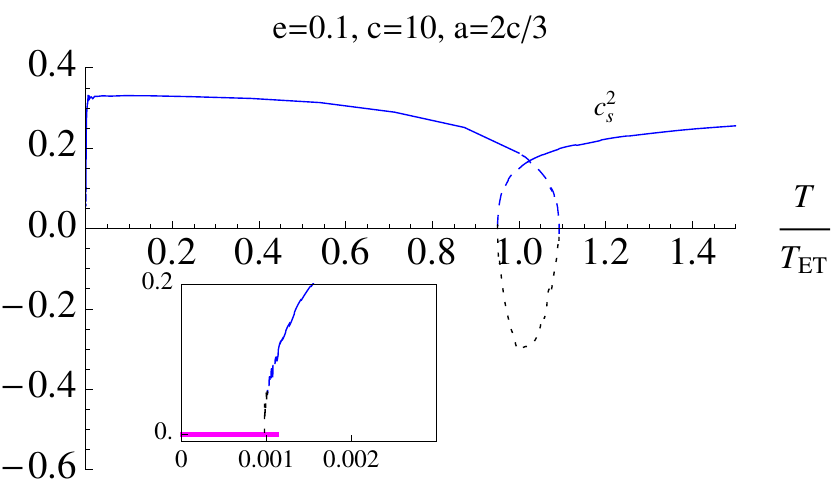}
\end{center}

\caption{\small Sound speed squared for the equation of state in Fig.\ref{eos_eneq0}.
The unstable region corresponds to $c_s^2<0$. $T$ approaches the conformal limit $1/3$ for
$T\gg\Thigh$ and also near $\Tlow$. Ultimately, as shown in
the inset, $c_s^2$ drops there to zero.
}
\la{cs_eneq0}
\end{figure}

Fig.~\ref{cs_eneq0} shows the sound speed squared computed for the equation of state
in Fig.~\ref{eos_eneq0}. Finally,
it is again of interest to compare the input beta function \nr{betafn}
with that what one obtains
by inserting numerically computed solutions for $\lambda(z),\,b(z)$ to \nr{crucial}.
The outcome for this $e\not=0$ case is shown in the left panel of Fig.\ref{betasene0}. For finite $T$
the computed beta functions terminate at some $\lambda$, when $T$ is decreased, the
computed beta functions approximate the input one better and better. However, there
is one small but very important difference: the input beta function \nr{betafn}
is parallel to the confinement line $-\fra32\lambda$ at large $\lambda$, but, due to
the added logarithmic term in \nr{Vansatz}, the output beta function crosses the
confinement line at the quasiconformal $\to$ confining transition. This explicitly
shows the role played by this logarithmic term. The right panel of Fig.\ref{betasene0} shows
the evolution of the coupling corresponding to the low $T$ output beta function of the left panel.

\begin{figure}[!tb]

\includegraphics[width=0.52\textwidth]{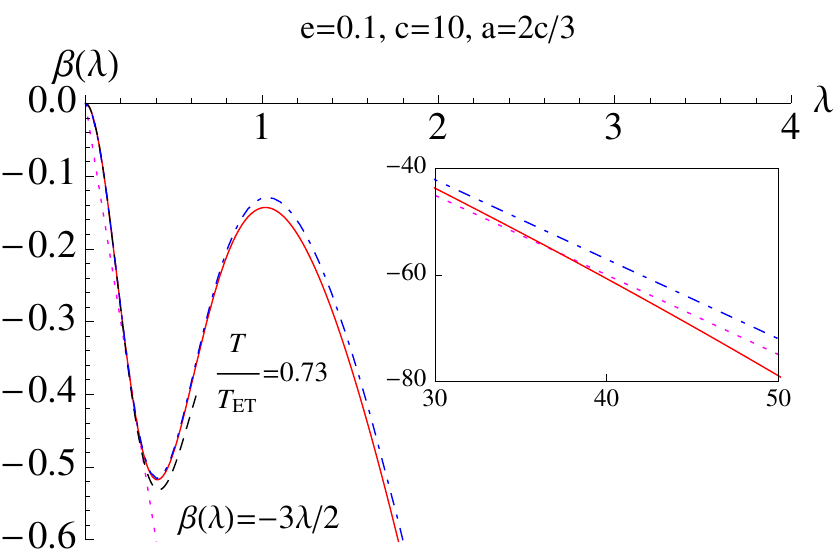}
\hfill 
\includegraphics[width=0.44\textwidth]{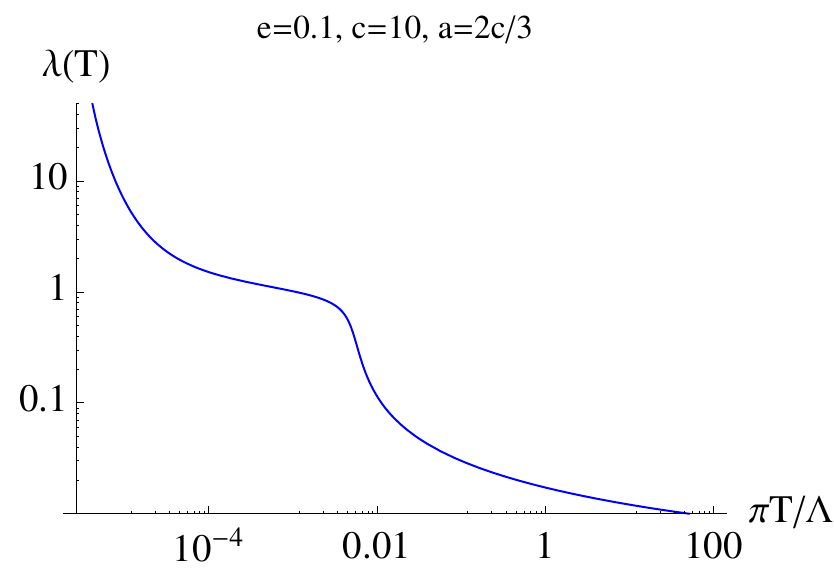}


\caption{\small Left: The input beta function (parametrisation \nr{betafn}, dotdashed line)
and the output beta function, computed from \nr{crucial} from the numerical solution at
$T=0.73\Thigh$ (dashed black line) and at the very low $T$ value $T=\Tlow$ (continuous
line),
for parameter values shown in the figure. The inset shows how the output beta function
at $T=\Tlow$ crosses the (dotted) confinement line \cite{kiri2} $-\fra32\lambda$
at the position of the quasi conformal $\to$ confined transition.
Right: Evolution of the coupling corresponding to the low $T$ output beta function. The $y$
axis here is the same as the $x$ axis of Fig.~\ref{Tlahfig}.}
\la{betasene0}
\end{figure}

To outline the full phase diagram of this theory, we shall first choose fixed values
in the first order transition region for the parameters $c$ and $a$. How the phase
diagram then depends on the parameter $e$ which controls the departure from the limit of
exactly infrared conformal theory is shown in Fig. \ref{phaseD}. The phase
structure along the $e=0$ axis corresponds to the results of Sec. \ref{IRFP}.
At the origin, $T=e=0$, the line of first order transitions, $\Tlow(e)$,
terminates at a second order quantum
transition, in which the vacuum theory becomes infrared conformal.
At finite values
of  $e\lsim 0.5$ there are, as a function of temperature, three phases corresponding to
confined,  quasi-conformal and deconfined matter. When $e$ increases, the upper transition
temperature $\Thigh(e)$ decreases slowly, while the lower transition temperature $\Tlow(e)$
increases rapidly. At some critical $e_c$ these lines merge and the quasi-
conformal phase disappears; for $e>e_c$
the phase structure will consist of a single transition line
separating deconfined and confined phases. In the case depicted in the figure
all transition lines correspond to a
first order transition and hence the intersection is a triple point.

Full phase diagram would require adding more axes corresponding to $c$ and $a$.
The inset in Fig.~\ref{phaseD} shows for what values of $c$ the transition changes
from a crossover to a 1st order one, for various $e$.

\begin{figure}
\begin{center}
\includegraphics[width=.6\textwidth]{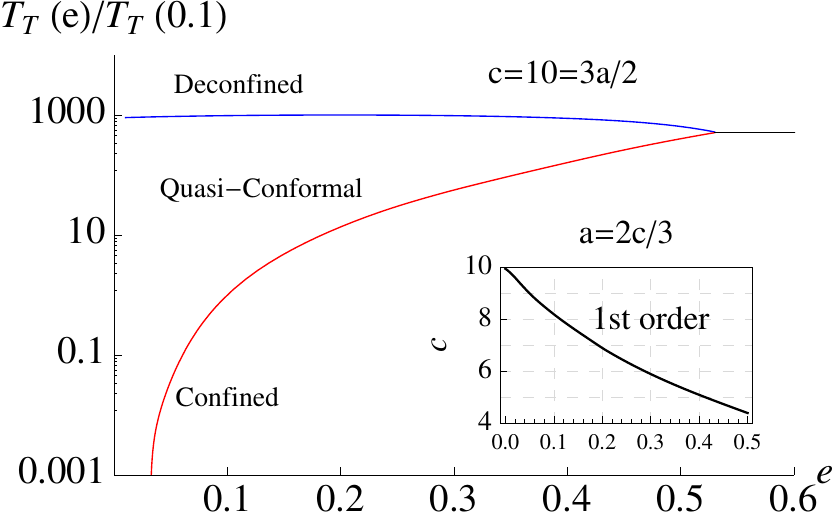}
\caption{The $(T,e)$ phase diagram for $c=3a/2=10$. The upper curve shows $\Thigh(e)$, the lower
$\Tlow(e)$, both normalised by $\Tlow(0.1)$. The inset shows for what values of $c$ for various $e$
the upper ET transition is of first order, for $c$ below the line the transition is a cross-over.
The figure is qualitative in the sense that
$\Thigh$ and $\Tlow$ have been approximated by the values of the minima of $T(\lambda_h)$.
The line to the right of the triple point is the deconfinement transition line.
}
\label{phaseD}
\end{center}
\end{figure}

\section{Discussion}
\label{disc}

Let us then discuss the implications of our results for lattice studies and cosmology.
On the lattice one has only the generic TC gauge theory, in cosmology all the
standard model degrees of freedom are involved.

In contrast to our
approach here, where no microscopic dynamics is assumed but only the features present in the
beta function of the theory affect the outcome, on the lattice one of course studies a specific SU($N$) gauge
theory with some number of matter fields transforming under a given representation of the gauge group.
In such theories, for given $N$ and fermion representation, the departure from conformal behavior
is controlled by the number of flavors. Generally then, the phase diagram in these cases is expected to be
similar to the one in Fig. \ref{phaseD} via identification $e\sim N_{f,\rmi{crit}}-N_f$,
where $N_{f,\rmi{crit}}$ denotes the
critical number of flavors at which the SU($N_c$) gauge theory under consideration develops
an infrared stable fixed point. As the number of flavors is decreased ($e$ increases) the
theory first becomes walking and then the conformal behavior vanishes. Lattice investigations
of the $(T,N_f)$ phase diagram in any theory shown to possess an infrared stable fixed point
at zero temperature for some $N_{f,\rmi{crit}}$ would therefore provide more insight into the phases
we have discussed in this paper.

Lattice studies of generic TC theories have so far only considered $T=0$. They involve all
the complications of lattice fermions and are extremely demanding. Extending them to $T>0$
with asymmetric lattices will be all the more so.

To discuss consequences for cosmology we need to couple the strong dynamics with the SM fields. The
possibilities are different for the cases of theories with
walking coupling or an infrared stable fixed point, so
we discuss them separately.

For the walking coupling the natural phenomenology framework is provided by TC. Then the scale
$\lambdalow$ is identified with $v_\rmi{weak}\approx 246$ GeV
and, if we assume ETC dynamics, the scale $\lambdahigh$ can be identified with
$\lambdaETC$. At high energies, above the scale $\lambdaETC$, massless ETC gauge bosons mediate
interactions between SM fermions and technifermions, while below $\lambdaETC$ the ETC gauge bosons
are massive due to dynamical symmetry breaking and Higgs mechanism associated with the ETC
gauge symmetries. As a consequence, between the scales
$\lambdalow$ and $\lambdaETC$, the SM  fields couple to the Technicolor theory via effective
four-fermion operators
\be
\frac{g_\rmi{ETC}}{\lambdaETC^2}{\mathcal{O}}_\rmi{SM}{\mathcal{O}}_\rmi{TC}.
\label{etcoperator}
\ee
Here the operators ${\mathcal{O}}_\rmi{SM,TC}$ represent bilinear fermion operators constructed from SM
and TC fields.

At the scale $\lambdalow$ the TC dynamics result in the Higgs mechanism and the electroweak gauge bosons
obtain their masses. In finite temperature this corresponds to the electroweak phase transition at $\Tlow$.
While the precise phenomenological details depend on the underlying gauge theory dynamics, the general
features are as follows: The technifermions $Q_L$ and $Q_R$ are singlet under QCD but have usual
electroweak
quantum numbers. The electroweak symmetry is embedded into the global chiral symmetry of the
technifermions so that the spontaneous breaking of this global symmetry results in correct breaking of
SU$_L(2)\times$U$_Y$(1) into U$_\rmi{em}$(1). As the chiral symmetry of the techniquarks is spontaneously
broken, TC is confined into TC singlet technihadrons. Note that this is an essential
{\em{assumption}} in our approach: as we have described in detail, within the holographic framework we
can identify the onset of confinement on the level of the beta function. Then, for the TC dynamics to
operate as we imagine here, the underlying gauge dynamics have to be such that deconfinement and chiral
symmetry restoration intertwine.
Typically this is the case and the exception is the theory with adjoint fermions
\cite{Mocsy:2003qw}.

As a concrete model for TC dynamics one can therefore consider SU($N$) with $N_f\sim 4N$ as
obtained from the ladder approximation and which is also compatible with recent lattice results
\cite{Appelquist:2007hu}; however, see also \cite{Fodor:2009ff}.
With higher representations a possible model would be SU(3) with two
flavors in the sextet representation. For SU(2) or SU(3) gauge theories with two adjoint flavors the dynamics
is expected to be richer due to chiral symmetry restoration and deconfinement remaining as independent
phase transitions with critical temperatures possibly widely separated.

The case of IRFP is related to very different phenomenology. Here we have only one
scale, denote this by $\lambdalow$. Being exactly conformal in the infrared, the strongly coupled theory does
not involve formation of a chiral condensate and hence cannot be used to break electroweak symmetry.
Nevertheless, our results can in principle be applied to unparticle cosmology; see \cite{McDonald:2008uh}
for a brief review. Due to lack of knowledge on the form of the unparticle operators, precise value of the
dynamical scale $\lambdalow$ and the nature of the Higgs sector we do not pursue the detail of this framework
here.

Finally, we remark that the walking beta function \nr{betafn} is a special case in
a class of beta-functions parametrized as
\be
\beta(\lambda)=-\tilde{c}\lambda^2\frac{\lambda^2+\tilde{f}\lambda+\tilde{e}}
{1+a_1\lambda+a_2\lambda^2+a_3\lambda^3}
\ee
In terms of this parametrization the ansatz in (\ref{betafn}) corresponds to
$\tilde{c}=c$, $\tilde{f}=-2$, $\tilde{e}=1+e$, $a_3=a$ and $a_1=a_2=0$.
Different parameter values
lead to very similar results once the parameters are chosen so that the resulting
beta-function has the three distinct regions corresponding to different evolutions
of the coupling constant in regions separated by scales $\lambdalow$ and
$\lambdahigh$. Therefore, for our investigation, the simple three parameter
form (\ref{betafn}) is convenient and sufficient. We have experimented with different functional forms which
reproduce the essential features discussed in previous sections and lead to very similar results for
finite temperature phase diagrams.
We expect that the results we have obtained are generic for a walking-type beta-function.

\section{Conclusions}
\label{checkout}
We have, in this paper, studied the thermodynamics of a field theory with the beta function
\nr{betafn} containing a quasi conformal region in which the coupling varies very slowly,
walks. As a limiting case, a theory with an infrared fixed point is also obtained. The
basis of the computation is a bottom-up gravity dual with a metric ansatz and a dilaton.
The results in this paper can be said to be a very concrete and productive
application of gauge/gravity duality.

An essential property of the approach is that the details of the 4d boundary field theory
need not be known, all of its properties are compressed in the beta function. This is
reminiscent of the applications of gauge/gravity duality to condensed matter physics:
there also the boundary theory is not written down, only expectation values are
computed. The price
one pays is that the beta function contains a number of unspecified parameters.

A central assumption in our analysis is that the knowledge of the functional form of
$\beta(\lambda)$ and the framework of \cite{kiri3,kiri4,aks,ak} provides a reasonable
description of the thermodynamics of the underlying microscopic theory even though
fermions in these theories are essential. The proper inclusion of flavor
in various representations in the
gauge/gravity correspondence setting,
remains a challenge for theorists. 

\vspace{1cm}
{\it Acknowledgements}.
We thank M. Antola, K. Rummukainen and V. Suur-Uski for discussions
and F. Nitti for the well documented numerical code used in \cite{kiri4}.
JA thanks the Magnus Ehrnrooth foundation for financial support.

\end{document}